\documentclass[10pt, twocolumn, pre, aps, superscriptaddress, showpacs]{revtex4-1}
\usepackage{amsmath, graphicx, subfigure, tikz}
\begin{document}

\title{Metastable Potts Droplets}
    \author{E. Can Artun}
    \affiliation{Faculty of Engineering and Natural Sciences, Kadir Has University, Cibali, Istanbul 34083, Turkey}
    \author{A. Nihat Berker}
    \affiliation{Faculty of Engineering and Natural Sciences, Kadir Has University, Cibali, Istanbul 34083, Turkey}
    \affiliation{Department of Physics, Massachusetts Institute of Technology, Cambridge, Massachusetts 02139, USA}

\begin{abstract}

The existence and limits of metastable droplets have been calculated using finite-system renormalization-group theory, for $q$-state Potts models in spatial dimension $d=3$.  The dependence of the droplet critical sizes on magnetic field, temperature, and number of Potts states $q$ has been calculated.  The same method has also been used for the calculation of hysteresis loops across first-order phase transitions in these systems.  The hysteresis loop sizes and shapes have been deduced as a function of magnetic field, temperature, and number of Potts states $q$.  The uneven appearance of asymmetry in the hysteresis loop branches has been noted.  The method can be extended to criticality and phase transitions in metastable phases, such as in surface-adsorbed systems and water.

\end{abstract}
\maketitle

\section{Introduction: Non-Equilibrium Properties from an Equilibrium Calculation}

Recently equilibrium renormalization-group calculations have been simply extended to the calculation of the properties of metastable droplets of the non-equilibrium phase surviving inside the equilibrium thermodynamic phase \cite{ErenBerker}.  This method was illustrated with the Ising model in $d=3$ spatial dimensions.  The limiting droplet sizes have been determined as a function of temperature and magnetic field.  The critical magnetic fields, above which no metastable droplet can exist, have been calculated as a function of temperature.  The method consists in making a finite-system renormalization-group calculation of the magnetization \cite{BerkerOstlund} and matching the boundary counditions of the outermost layer of the droplet.  If this reverse magnetization sustains inside the droplet, the droplet exists and otherwise not, for the given droplet size.

\begin{figure}[ht!]
\centering
\includegraphics[scale=0.38]{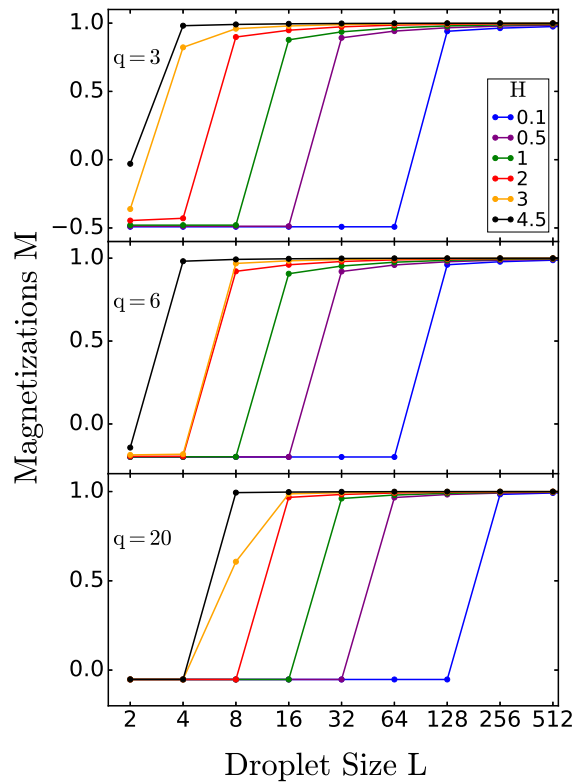}
\caption{Metastable droplet magnetizations $M = [\left<\delta(s_i,1)\right> - 1/q]/(1-1/q)$ as a function of droplet
size, at temperatures $T/T_C=J_C/J=0.25$. The droplet exists when the magnetization is negative. In each panel for each number $q$ of Potts states, the lines are for magnetic fields $H=0.1,0.5,1,2,3,4.5$ from right to left. The maximal droplet size is the average of the lengths at each end of the rise from negative to positive calculated magnetization.}
\end{figure}

\begin{figure}[ht!]
\centering
\includegraphics[scale=0.45]{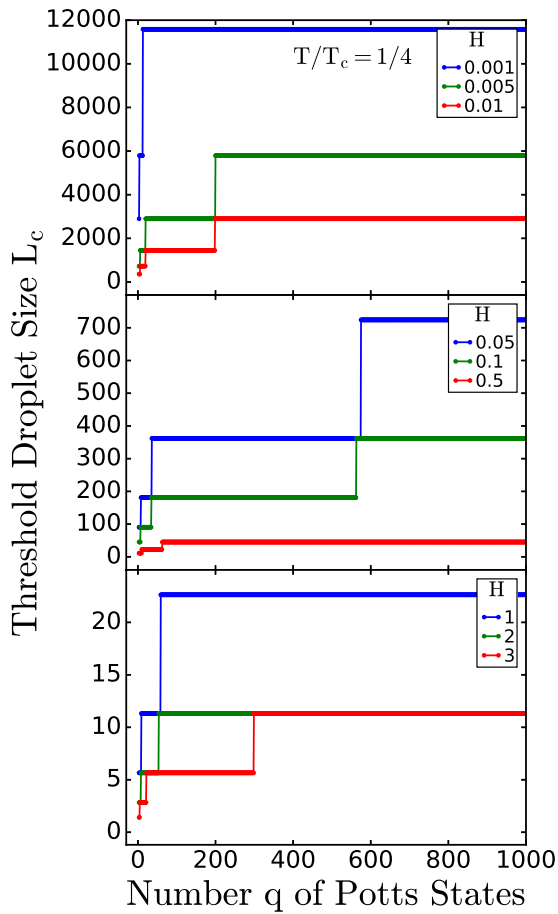}
\caption{Maximal droplet sizes as a function of the number $q$ of Potts states, for different magnetic fields, at temperatures $T/T_C=J_C/J=0.25$. The multi-stepped curves are, from top to bottom in each panel, for $H=0.001,0.005,0.01$ (top panel), $0.05,0.1,0.5$ (middle panel), $1,2,3$ (bottom panel).  Note from vertical axis values, the wide range of droplet sizes under different conditions. A trend is seen in the droplet size values, but not in the step occurrences.}
\end{figure}

In the present study, we have extended this work to $q$-state Potts models for arbitrary $q$ in $d=3$.  We determine the threshold droplet sizes as a function of the number of states $q$ and find changes even at high values of $q$, similarly to the equilibrium thermodynamic properties of the Potts models \cite{ArtunBerker}.  The method also naturally yields the calculation of hysteresis loops, which yields a large variety as a function of $q$, domain size, and temperature.

\begin{figure}[ht!]
\centering
\includegraphics[scale=0.45]{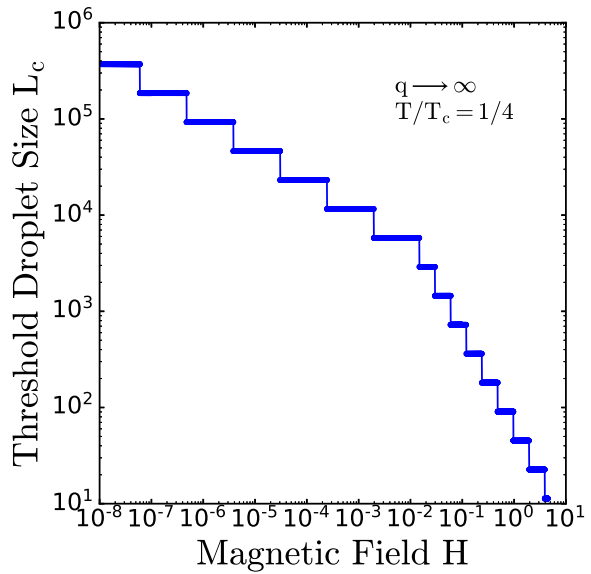}
\caption{Maximal droplet sizes as a function of magnetic field, at temperature $T/T_C=J_C/J=0.25$. As seen in Fig. 2, the maximal droplet size occurs at large $q$. A crossover in power-law behavior is clearly seen, from $L_C \sim H^{-0.99}$ at low $H$ to $L_C \sim H^{-0.33}$ at very low $H$.}
\end{figure}

\begin{figure}[ht!]
\centering
\includegraphics[scale=0.39]{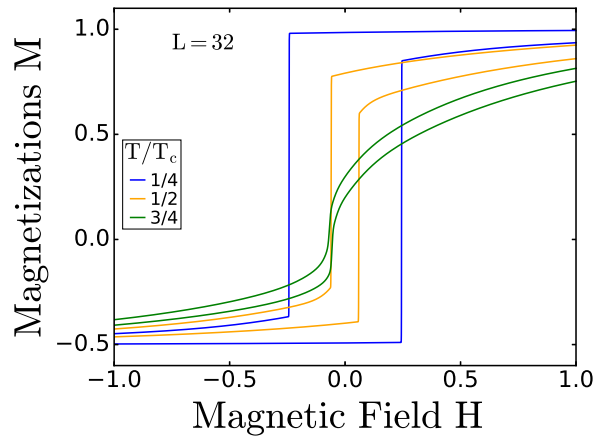}
\caption{Hysteresis loops for different temperatures, for $q = 3$.  From outer to inner, the loops are for temperatures $T/T_C=J_C/J=0.25, 0.50, 0.75$.  As the temperature approaches $T_C$, the hysteresis loops get narrower and the two branches composing the loop acquire curvature starting from the non-leading side with respect to scanning direction.}
\end{figure}

\begin{figure*}[ht!]
\centering
\includegraphics[scale=0.4]{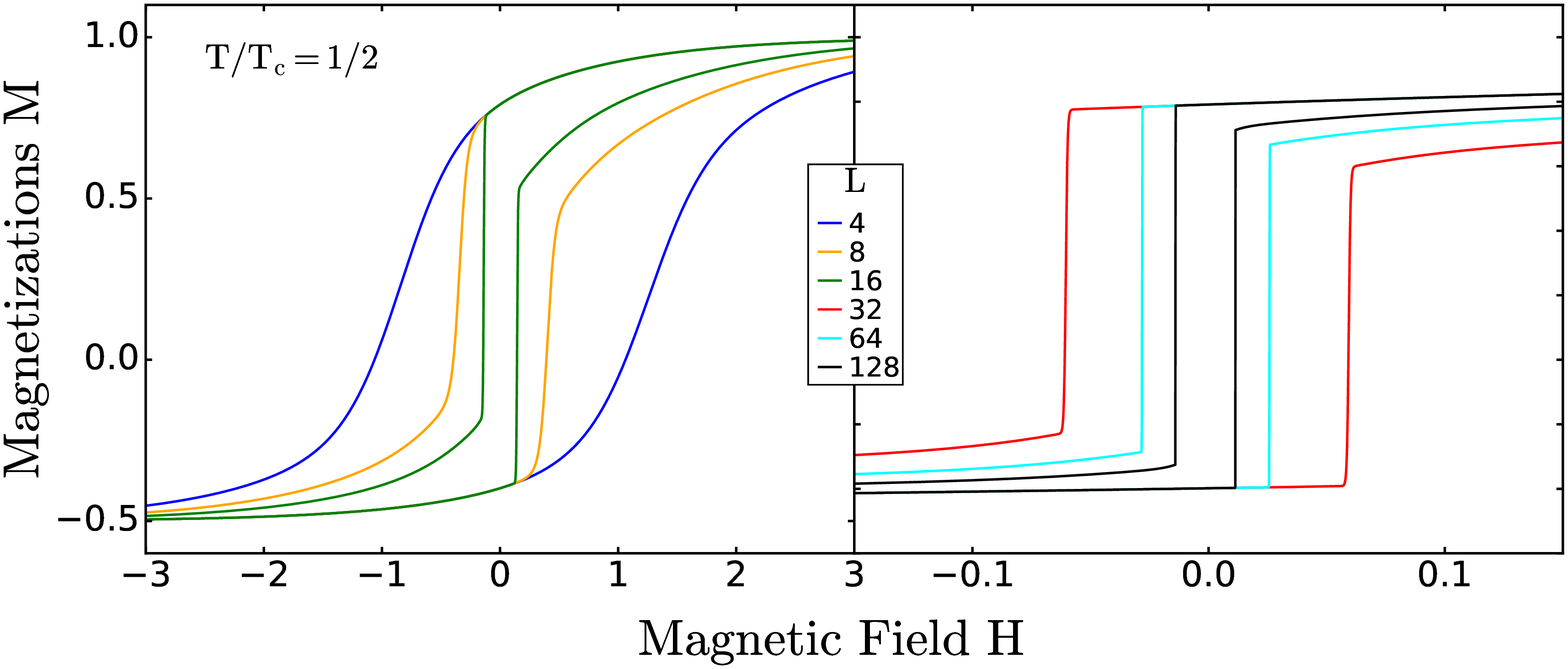}
\caption{Hysteresis loops for different sizes, for $q = 3$ and temperature $T/T_C=J_C/J=0.50$. From outer to inner, the loops are for sizes $L=4,8,16$ (on left) and $32,64,128$ (on right).  As the system size increases, the hysteresis loops get narrower and acquire
vertical edges.}
\end{figure*}

Our study introduces a generalization of the position-space renormalization-group methods - originally used for equilibrium thermodynamic phases - to metastable droplets with $q$ states.  Whereas traditional mean-field theories have historically yielded important generic intuition on the phase transition problem, position-space renormalization-group methods on the other hand can be taylored to distinctive system attributes, for example yielding experimental phase diagrams \cite{Surface1,Surface2,Surface3}, second-to-first-order phase transition changeovers by effective vacancies \cite{NienhuisPotts,AndelmanBerker}, scaling chaos in frustrated systems \cite{McKayChaos}, first-to-second-order phase transition changeovers by random magnetic magnetic fields \cite{HuiBerker,erratum}, etc.  An extension to non-equilibrium is certainly desirable.

\section{Model and Method}

The Potts models are defined by the Hamiltonian:
\begin{equation}
- \beta {\cal H} =
\sum_{\left<ij\right>} \{J[\delta(s_i,s_j)-1/q] + H [\delta(s_i,1)+\delta(s_j,1)]\}\,,
\end{equation}
where $\beta=1/k_{B}T$, at site $i$ the spin $s_{i}=1,2,...,q$ can
be in $q$ different states, the delta function $\delta(s_i,
s_j)=1(0)$ for $s_i=s_j (s_i\neq s_j)$, and $\langle ij \rangle$
denotes summation over all nearest-neighbor pairs of sites. We have used the traceless form of interaction in the first term of Eq.(1). Under renormalization-group, the Hamiltonian is conveniently expressed as
\begin{equation}
- \beta {\cal H} =  \sum_{\left<ij\right>} [E(s_i,s_j)+G] \, ,
\end{equation}
The last term in Eq.(2) is the additive constant that is unavoidably generated by the renormalization-group transformation and that is essential in the calculation of the thermodynamic densities, as seen below.  With no loss of generality, after each renormalization-group transformation, G is fixed so that the largest energy $E(s_i,s_j)_{max}$ of the spin-spin interaction is zero (and all other $E(s_i,s_j) < 0$). This formulation makes it possible to follow global renormalization-group trajectories, necessary for the calculation of densities for the point at the onset of the renormalization-group trajectory, without running into numerical overflow problems.

As the renormalization-group transformation, we use the Migdal-Kadanoff approximation \cite{Migdal,Kadanoff} with length rescaling factor $b=2$, which is also the exact transformation for a $d=3$ hierarchical lattice \cite{BerkerOstlund,Kaufman1,Kaufman2}.  (However, it will be seen below that our method is usable with any renormalization-group transformation.  We have used the Migdal-Kadanoff approximation here, as it is easily implemented and has been quite successful in a variety of systems.) This transformation consists in a bond moving followed by a decimation, giving the renormalization-group recursion relations.  The transformation is very simply expressed in terms of the transfer matrix $\mathbf{T(s_i,s_j)} = e^{E(s_i,s_j)}$:  Bond moving consists of taking the power of each element of the transfer matrix, $\widetilde{T}(s_i,s_j) = [T(s_i,s_j)]^{b^{d-1}}$.  Decimation consists of matrix multiplication,
\begin{equation}
\mathbf{T'} = \mathbf{\widetilde{T}\cdot\widetilde{T}} \, e^{\widetilde{G}} \, ,
\end{equation}
where $\widetilde{G}$ is chosen so that the largest energy $E'(s_i,s_j)_{max}$ of the spin-spin interaction is zero  as explained above. The recursion relation for the additive constant is then $G' = b^d G + \widetilde{G}$.  The primes denote the quantities of the renormalized system.

The densities are calculated by the density recursion relation of the renormalization-group transformation, $\mathbf{M} = b^{-d}\mathbf{M'\cdot R}$, where $\mathbf{M} = [1,\left<\delta(s_i,m)\delta(s_j,n)\right>]$ are the densities conjugate to the energies $\mathbf{K} = [G, E(m,n)]$, where $m,n$ span the Potts states $(1,...,q)$ and $n=m$ is not included in $\mathbf{K}$, since in our calculation these correspond to the leading energies and are always set to zero as explained above, by fixing $\widetilde{G}$.  The recursion matrix is $\mathbf{R} = \mathbf{\partial K' / \partial K}$.  By multiply self-imbedding this density recursion relation, $\mathbf{M^{(0)}} = b^{-d}\mathbf{M^{(n)}\cdot R^{(n)}\cdot ... \cdot R^{(1)}}$, where $\mathbf{M^{(m)}}$ are the densities at the energies $\mathbf{K^{(m)}}$ reached after the $(m)$th renormalization-group iteration.  Our calculation of the densities $\mathbf{M^{(0)}}$ is done by using the droplet boundary condition for $\mathbf{M^{(n)}}$ where $L = b^n$ is the size of the would-be droplet.  We perform our metastable droplet calculations for $H>0$ in Eq.(1), so that the magnetization $M = [\left<\delta(s_i,1)\right> - 1/q]/(1-1/q)$ is positive (negative) in the equilibrium thermodynamic phase (metastable phase) and $\left<\delta(s_i,m\neq 1)\delta(s_j,m)\right> = 1$ is the metastable droplet boundary condition.

\section{Results: Metastable Droplets}

The calculated metastable droplet magnetizations $M = [\left<\delta(s_i,1)\right> - 1/q]/(1-1/q)$ as a function of droplet size $L$ are given in Fig. 1, for temperatures $T/T_C=J_C/J=0.25$, where $1/J_C$ is the equilibrium critical temperature, obtained \cite{BerkerOstlund,ArtunBerker} for each $q$ from the solution of $x = (x^8+q-1)/(2x^4+q-2)$, where $x=e^{J_C}$.  The droplet exists when the magnetization is negative. As explained at the end of Sec. II, positive magnetization means predominantly being in the one state favored by the magnetic field and negative magnetization means predominantly being in any one of the $q-1$ states not favored by the magnetic field. The magnetic field in Eq. (1) is defined so as to favor one of the Potts states. Thus all possible Potts state droplets are covered in our work. In each panel of Fig. 1 for each number $q$ of Potts states, the lines are for magnetic fields $H=0.1,0.5,1,2,3,4.5$ from right to left.  The magnetization discontinuity occurs at the maximal droplet size $L_C$ for each $q$ and $H$.  Higher magnetic field $H$ energetically favors the equilibrium thermodynamic phase, moves the system away from $H=0$ where the metastable phase also becomes a stable thermodynamic phase, and represses the metastable droplet.

The thus calculated maximal droplet sizes as a function of the number $q$ of Potts states for different magnetic fields, at temperatures $T/T_C=J_C/J=0.25$, are given in Fig. 2. From vertical axis values, a wide range of droplet sizes under different conditions is seen.  The multi-stepped curves show changes even at unusually high values of $q$, namely in the hundreds.  This is akin to the equilibrium properties of the Potts models, where the phase transition temperature does not saturate as a function of $q$, unlike the similar clock models \cite{ArtunBerker}.

Maximal droplet sizes as a function of magnetic field, at temperature $T/T_C=J_C/J=0.25$, are given in Fig. 3. As seen in Fig. 2, the maximal droplet size occurs at large $q$. A crossover in power-law behavior is clearly seen, from $L_C \sim H^{-0.99}$ at low $H$ to $L_C \sim H^{-0.33}$ at very low $H$.

\section{Results: Hysteresis Loops}

Another common non-equilibrium occurrences are hysteresis loops, where, in scanning across a first-order phase transition, the system retains the memory of previous steps, via pinned spins at the boundaries of microdomains or at impurities, or slow dynamics.  Our method is easily applicable to this phenomenon.  In performing our density calculation for a finite microdomain, we keep the boundary condition pinned at the $q=1$ phase when scanning down in magnetic field and at the $q\neq 1$ phases when scanning up in magnetic field.  Thus, in all of our results seen in Figs. 4-6, the upper (lower) branch of the hysteresis loop is obtained for scanning down (up) in magnetic field.

\begin{figure}[ht!]
\centering
\includegraphics[scale=0.4]{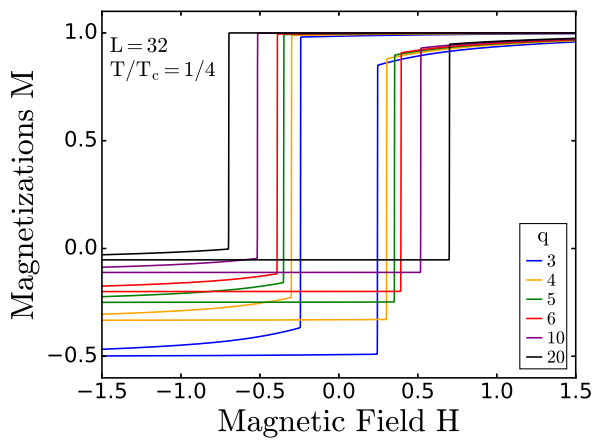}
\caption{Hysteresis loops for different number of states $q$, for temperatures $T/T_C = 1/4$ and $L=32$. From outer to inner, the loops are for $q= 20,10,6,5,4,3$. As the number of states decreases, the hysteresis loops get narrower and acquire curvature at the non-leading side with respect to scanning direction of each branch.}
\end{figure}

Hysteresis loops for different temperatures, for $q = 3$, are shown in Fig. 4.  From outer to inner, the loops are for temperatures $T/T_C=J_C/J=0.25, 0.50, 0.75$.  As the temperature approaches $T_C$, the hysteresis loops get narrower and the two branches composing the loop acquire curvature starting from the non-leading side with respect to the scanning direction.  Hysteresis loops for different sizes, for $q = 3$ and temperature $T/T_C=J_C/J=0.50$, are given in Fig. 5. As the system size increases, the hysteresis loops get narrower and acquire vertical edges.  At infinite system size, the single discontinuous curve of the equilibrium first-order phase transition obtains. Hysteresis loops for different number of states $q$, for temperatures $T/T_C = 1/4$ and $L=32$, are given in Fig. 6. As the number of states decreases, the hysteresis loops get narrower and acquire curvature at the non-leading side of each branch.

\section{Conclusion: Metastable Criticality}

It is seen that metastable phase droplet properties can readily be calculated, using finite-system renormalization-group theory, for a variety of systems.  Furthermore, critical phenomena and phase transitions in metastable phases have been discussed, in the past, for important physical systems, such as surface-adsorbed systems \cite{Griffiths1,Griffiths2} and water \cite{Stanley1,Stanley2}.  Our method can be applied to study such metastable criticality and phase transitions.  The extension of our method to $q$-state Potts models thus increases the range of possible experimental applicability \cite{Surface1,Surface2,Surface3}.

In going from the metastable Ising droplets \cite{ErenBerker} to the metastable Potts droplets of the current work, we have found a very wide range of metastable droplet sizes under different number $q$ of Potts states  and magnetic field conditions. For a fixed number $q$ of Potts states, the two branches composing the hysteresis loops acquire curvature as the temperature is increased towards the critical temperature, starting from the non-leading side with respect to the scanning direction.  An identical effect occurs for a fixed temperature, the two branches composing the hysteresis loops acquiring curvature starting from the non-leading side with respect to the scanning direction, as the number $q$ of Potts states is increased.  It was previously shown, within the context of antiferromagnetic Potts models, that increasing temperature and increasing $q$ have similar entropic effects \cite{BerkerKadanoff1,BerkerKadanoff2}.

Finally, we hope that numerical experiments, namely computer simulations, will check the maximum metastable droplet and hysteresis loop phenomena predicted by our theory.  An important extension of our theory would be to the dynamics of metastable droplet disappearance. Thus, the droplet disappearance, either from the interior or from the periphery, would be distinguished by its effects.

\begin{acknowledgments}
Support by the Kadir Has University Doctoral Studies Scholarship
Fund and by the Academy of Sciences of Turkey (T\"UBA) is gratefully
acknowledged.
\end{acknowledgments}

\end{document}